\documentclass[aps,prx,floatfix,twocolumn,superscriptaddress,showpacs,epsf]{revtex4-2}

\usepackage{amsmath,amssymb,amsfonts}
\usepackage{mathrsfs}
\usepackage{epstopdf}
\usepackage{epsfig} 
\usepackage{graphicx}
\usepackage{subfigure}
\usepackage{import}
\usepackage{color}
\usepackage{url}
\usepackage{bm}\let\vec\bm
\usepackage{changes}
\usepackage{natbib}
\usepackage{braket}

\allowdisplaybreaks

\usepackage[colorlinks=true,linkcolor=blue,citecolor=red]{hyperref}

\newif\ifshowcomments\showcommentstrue

\usepackage{changes}

\makeatletter
\renewcommand\tableofcontents{\@starttoc{toc}}
\makeatother
\def\beq{\begin{equation}}
\def\eeq{\end{equation}}

\def\vk{\vec{k}}

\def\ek{\varepsilon_{\mathbf k}}

\newcommand{\ee}{\varepsilon}

\newcommand\ims{\mathrm{Im}\Sigma}
\newcommand\res{\mathrm{Re}\Sigma}

\def\taup{\tau_{+}}
\def\taum{\tau_{-}}

\newcommand{\Se}{S}

\newcommand{\ph}{\Phi_0}
\newcommand{\php}{\Phi^{\prime}_0}
\newcommand{\gamel}{\gamma}
\newcommand{\gamin}{\Gamma_{\mathrm{in}}}
\newcommand{\ombar}{x}
\newcommand{\gm}{g_{-}}
\newcommand{\cm}{c_{-}}

\newcommand{\prefZ}{c_Z}
\newcommand{\el}{\mathrm{el}}
\newcommand{\inel}{\mathrm{in}}
\newcommand{\ts}{\theta}

\newcommand{\taubar}{\overline{\tau}}
\usepackage{fancyhdr}
\begin{document}
\author{Antoine Georges}
\affiliation{Coll{\`e}ge de France, 11 place Marcelin Berthelot, 75005 Paris, France}
\affiliation{Center for Computational Quantum Physics, Flatiron Institute, New York, NY 10010 USA}
\affiliation{CPHT, CNRS, Ecole Polytechnique, IP Paris, F-91128 Palaiseau, France}
\affiliation{DQMP, Universit{\'e} de Gen{\`e}ve, 24 quai Ernest Ansermet, CH-1211 Gen{\`e}ve, Suisse}
\author{Jernej Mravlje}
\affiliation{Department of Theoretical Physics, Institute Jo\v{z}ef Stefan, Jamova 39, SI-1001 Ljubljana, Slovenia. } 

\title{Skewed Non-Fermi Liquids and the Seebeck Effect}

\begin{abstract}
We consider non-Fermi liquids in which the inelastic scattering rate 
has an intrinsic particle-hole asymmetry and obeys $\omega/T$ scaling.  
We show that, in contrast to Fermi liquids, this asymmetry influences the 
low-temperature behaviour of the thermopower even when the impurity scattering dominates. 
Implications for the unconventional sign and temperature dependence of the thermopower in cuprates 
in the strange metal (Planckian) regime are emphasized.
  \end{abstract}
\maketitle
\section{Introduction}
Besides its relevance to thermoelectricity, the Seebeck effect provides invaluable insights 
into the fundamental physics of materials\cite{goldsmid_book,behnia_book,AG_lectures}. 
The Seebeck coefficient $S$ (thermopower) is 
sensitive to the 
balance between hole-like and electron-like excitations. 
It is negative when electrons dominate, positive when holes dominate 
and vanishes when particle/hole symmetry holds. 
In many cases, the particle-hole asymmetry originates in the bandstructure of the material: it is controlled 
by the number (density of states) and velocities of the two types of excitations. 
However, it has been recognized that another source of asymmetry may also influence the Seebeck coefficient: 
that of the lifetime (scattering rate) of these excitations. 
Although this has been discussed 
theoretically~\cite{Robinson_1967,haule_2009,Gweon_2011,Shastry_2012,Deng_2013,Xu_2013,Zitko_2013,shahbazi16,Zitko_2018}  
and put forward as a possible mechanism 
for materials in which the sign of $S$ 
is found to be opposite to that predicted by bandstructure~\cite{Robinson_1967,shabazi16,Xu_2014,Xu_2020}, it has received comparatively less attention. 
One of the reasons is that, as detailed below, 
the particle-hole asymmetry of the inelastic scattering rate does not influence $S$ at  
low-temperature for metals obeying Fermi liquid theory when impurity scattering is also present. 

In this article, we show that the situation is entirely different in correlated metals which do not obey Fermi liquid theory. 
We consider a family of non-Fermi liquids in which the inelastic (electron-electron) scattering rate $\gamin$ obeys $\omega/T$ scaling 
(with $\omega$ the energy of an excitation counted from Fermi level). 
We demonstrate that in `skewed' non-Fermi liquids where the scaling function 
has an odd frequency component, this particle-hole asymmetry affects the low-$T$ behaviour of the Seebeck 
coefficient down to $T=0$, even in the presence of impurity scattering. 
This is an unexpected finding because the impurity scattering is temperature independent, whereas the electron-electron scattering diminishes upon cooling down 
and vanishes at $T=0$.
The sign of $S$ can be reversed in comparison to that expected from bandstructure. 
The case of a `Planckian' metal~\cite{Zaanen04,Marel2003,Bruin_2013,Hartnoll_2014,Legros19,Grissonnanche_2020,Varma_2020} 
with $\gamin\propto \omega,T$ 
turns out to be particularly interesting. 
In that case, $S/T$ ultimately diverges logarithmically at low temperature. However, the temperature dependence and sign 
of $S/T$ over an extended temperature range is strongly affected by the particle-hole asymmetry of $\gamin$.  
As discussed below, 
this may be relevant to the understanding of the Seebeck coefficient of 
cuprate superconductors, especially close to the critical doping where the pseudo-gap opens and a 
logarithmic dependence of the specific heat is observed~\cite{Michon_2019}. 

The paper is organized as follows: In Sec.~\ref{sec:meth} we 
describe the Kubo-Boltzmann formalism used in our calculations of the Seebeck coefficients, 
taking explicit account of the skeweness (particle-hole asymmetry) of the inelastic scattering rate. 
In Sec.~\ref{sec:fl_vs_nfl} we describe our main results that reveal the unusual effect of skeweness in non-Fermi 
liquids and contrast it to the more moderate behavior in Fermi liquids. In Sec.~\ref{sec:disc} we discuss the implication of our findings for experiments. Technical details are delegated to appendices: In Appendix~\ref{app:kubo} we give a derivation of the  transport equations starting from the Kubo formalism, in Appendix~\ref{app:skewed} we discuss the $\omega/T$ scaling properties of the considered class of non-Fermi liquids, and in Appendix~\ref{app:T} we elaborate on the behavior of the Seebeck coefficient at higher temperatures.

\section{Seebeck coefficient and particle-hole asymmetric inelastic scattering}
\label{sec:meth}
We recall that in metals, for non-interacting electrons and in the presence of elastic scattering only, 
the Seebeck coefficient 
at low temperature $T$ 
is given by: 
\begin{equation}
    \frac{S}{T} = - \frac{k_B}{e} \frac{\pi^2}{3} \frac{\php}{\ph}
\end{equation}
This expression involves the transport function $\Phi(\ee)=2\int d^dk/(2\pi)^d\, v_{\vk}^2\, \delta(\ee-\ek)$ at 
the Fermi level $\ph=\Phi(\ee_F)$ and its derivative with respect to energy $\php=\Phi^\prime(\ee_F)$. 
$v_{\vk}=\left(\nabla_{\vk}\ek\right)_\alpha/\hbar$ denotes the electron velocity in the direction $\alpha=x,y,z$ considered 
(we set $\hbar=k_B=1$ in most of the following).  
In this simplest description, $S/T$ does not depend on the magnitude of the scattering rate and 
its sign is determined by the particle-hole asymmetry of the bandstructure encoded in the transport function: 
$S/T>0$ for a hole-like bandstructure ($\php<0$) and $S/T<0$ for an electron-like bandstructure ($\php>0$)

We now consider the effect of both elastic and inelastic scattering (that is, the impurity and the electron-electron scattering, respectively), writing the total scattering rate as: 
\begin{equation}
\Gamma(T,\omega)\,=\,\gamel + \gamin(T,\omega)  
\label{eq:elastic}
\end{equation}
It is convenient for our purpose to decompose the scattering {\it time} into components which are even and odd in frequency: 
\begin{equation}
    \tau_{\pm}(T,\omega) = \frac{1}{2}\left[\frac{1}{\Gamma(T,\omega)}\pm\frac{1}{\Gamma(T,-\omega)} \right]
\end{equation}
%
Note that we assume 
that the elastic scattering rate $\gamma$ is isotropic and that the inelastic scattering rate only depends on frequency and not on momentum (i.e. does not vary along the Fermi surface). 
The isotropy assumption allows us to keep the discussion simple and is sufficient 
to reveal the main effects 
that we wish to emphasize. 
Likewise, in our calculation of transport we do not take into account vertex corrections. 
This simplification is exact in models in which both the 
self-energy $\Sigma$ and the many-particle vertex are local (momentum-independent)
~\cite{Khurana_1990}. We thus take 
$\gamin(T,\omega)=-2\mathrm{Im}\Sigma(\omega+i0^+,T)$ 
(also neglecting possible distinctions between current and energy relaxation rates~\cite{lavasani_2019,shouhang_2020}). 

Starting from the Kubo formula, one can derive the following expression for the Seebeck coefficient in the low-temperature regime~\cite{note_lowT}, as detailed in Appendix~\ref{app:kubo} : 
\beq
\Se = - \frac{k_B}{e} \frac{I_1(T)}{I_0(T)}
\label{eq:Seebeck}
\eeq
in which: 
\begin{eqnarray}
I_1(T)&=&\frac{T}{Z(T)}\frac{\php}{\ph}\, \langle \ombar^2\taup\rangle + \langle \ombar\taum\rangle
\label{eq:I1}
\\
I_0(T)&=&\langle \taup \rangle + \frac{T}{Z(T)}\frac{\php}{\ph}\, \langle x\taum\rangle
\label{eq:I0}
\end{eqnarray}
In these expressions, the frequency dependence of the scattering rates is expressed in terms of the scaling variable 
$\ombar=\omega/T$: $\tau_{\pm}=\tau_{\pm}(T,xT)$ and we use the notation:
$\langle F(x) \rangle\,\equiv\,\int_{-\infty}^{+\infty} dx F(x)/4\cosh^2\frac{x}{2}$. 
$Z$ denotes the effective mass renormalisation, which for a local theory is related to 
the real part of the self-energy by: 
$1/Z(T,\omega)=1+\left[\mathrm{Re}\Sigma(0)-\mathrm{Re}\Sigma(\omega)\right]/\omega$. 
Strictly speaking, the frequency dependence of $Z$ has to be kept in the integrals entering Eq.~(\ref{eq:I1},\ref{eq:I0}). 
As we discuss in Appendix~\ref{app:kubo}, neglecting this effect is actually 
a good approximation, and we use it here to simplify the discussion. 
In a local Fermi liquid, $Z(T)$ coincides with the quasiparticle spectral weight and reaches a finite value at $T=0$, 
while in a non-Fermi liquid $Z(T)$ may vanish as $T\rightarrow 0$~\cite{note_Z}. 
%
In the absence of interactions ($Z=1$) and for elastic scattering only ($\taup=1/\gamel, \taum=0$) we recover 
from Eqs.~(\ref{eq:Seebeck},\ref{eq:I1},\ref{eq:I0}) the 
simple expression Eq.~(\ref{eq:elastic}) ($\langle 1\rangle =1, \langle x^2\rangle = \pi^2/3$). 
The effect of a possible temperature dependence of $Z(T)$ on both the Seebeck coefficient and the specific heat/entropy, 
for example near a quantum critical point, has been previously discussed in the literature, see e.g. Ref.~\cite{Paul_2001}. 
In the present work, we focus on the effect of a particle hole asymmetry of the scattering rate, i.e on the terms 
involving $\langle x\taum\rangle$ in Eqs.~(\ref{eq:I1},\ref{eq:I0}).

\section{Fermi liquids vs. non-Fermi liquids}
\label{sec:fl_vs_nfl}
\subsection{Fermi liquid}
\begin{figure}
 \begin{center}
   \includegraphics[width=1.0\columnwidth,keepaspectratio]{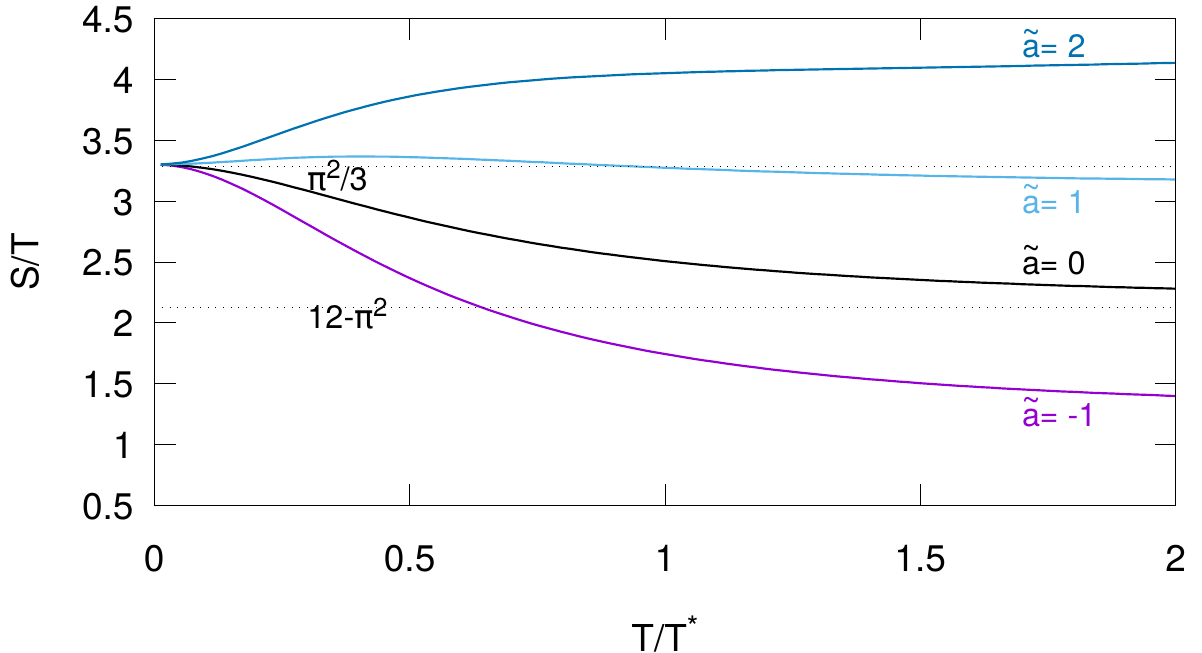}
   \end{center}
   \caption{\label{fig:fl}Fermi liquid: $S/T$ in units of $|k_B/e\cdot\php/\ph|$ vs. $T/T^*$ for several values of the 
   dimensionless particle-hole asymmetry parameters $\widetilde{a}\equiv a\ph/\php$ and taking $a=b$ (see text). 
   This illustrates the crossover between the low-$T$ elastic-dominated regime which does not depend on asymmetry and  
   the higher-$T$ inelastic-dominated one in which the asymmetry contributes. 
   A hole-like band contribution $\php/\ph <0$ is considered here.}
 \end{figure}
We first consider a (local) Fermi liquid with an inelastic scattering rate:  
\begin{equation}
    \gamin(T,\omega)=\lambda\left[
    \omega^2+(\pi T)^2 + a\omega^3+b\omega T^2\right] + \cdots
\end{equation}
The key point is that the odd part of the scattering rate scales as $\omega^3,\omega T^2\sim xT^3$ 
and hence is {\it subdominant} as compared to the even-frequency part: 
the inelastic scattering rate of conventional Fermi liquids is asymptoticallly particle-hole 
symmetric at low energy. 
Adding the elastic scattering rate, we see that there are two regimes. 
In the `elastic regime' $\gamma$ dominates over the inelastic scattering rate: this holds for 
$\gamma\gtrsim \lambda(\pi T)^2$ or alternatively for $T\lesssim T^*$ with 
$T^*\sim (\gamel/\pi^2\lambda)^{1/2}$ a crossover temperature. 
In the `inelastic' regime ($\gamma\lesssim \lambda(\pi T)^2$ or $T\gtrsim T^*$), inelastic scattering 
dominates~\cite{note_lowT}.
%
Let us consider first the `inelastic' regime, in which  
$\taup$ is of order $1/T^2$, and $\taum$ of order $1/T$. 
Hence, in the denominator $I_0$ of Eq.~(\ref{eq:Seebeck}), $\braket{\taup}\sim 1/T^2$ dominates over $T\braket{x\taum}\sim \mathrm{const.}$. 
In contrast, in the numerator $I_1$, the odd term $\braket{x\taum}$ has the same $1/T$ temperature dependence as  
the even one $T\braket{x^2\taup}$. 
Taking into account that, in a Fermi liquid, the effective mass enhancement 
$1/Z=m^*/m$ reaches a constant at low temperature, we obtain in the inelastic limit $T\gg T^*$: 
\begin{equation}
    \frac{S}{T}\bigg|_{\inel}^{FL}\simeq -\frac{k_B}{e} \left[(12-\pi^2)\frac{1}{Z} \frac{\php}{\ph} 
    - \frac{12}{\pi^4} (c_a a+ c_b b)\right]
    \label{eq:FL-HiT}
\end{equation}
with $c_a=\braket{x^4/(1+x^2/\pi^2)^2}\simeq 6.51$, $c_b=\braket{x^2/(1+x^2/\pi^2)^2}\simeq 1.09$. 
Remarkably, the odd-frequency terms of $\gamin$ directly contribute to $S$ in this limit, on equal footing with the bandstructure term. 
This was, to our knowledge, first emphasized in Ref.~\cite{haule_2009} in which expression Eq.~(\ref{eq:FL-HiT}) was derived, 
and further discussed in Refs~\cite{Gweon_2011,Shastry_2012,Deng_2013,Xu_2013,Zitko_2013,Zitko_2018}. 
We also note that the prefactor of the  bandstructure term is modified (from $\pi^2/3\simeq 3.29$ to $12-\pi^2\simeq 2.13$) 
as compared to the elastic (low-$T$) limit. 

However, in the low-$T$ `elastic' limit ($\gamel\gg\lambda\pi^2T^2$ or $T\ll T^*$), this interesting effect disappears. 
Indeed, $\taup\sim 1/\gamel$ is constant in this limit, while $\taum$ vanishes as $T^3$, leading to: 
\begin{equation}
\frac{S}{T}\bigg|_{\el}^{FL}\simeq -\frac{k_B}{e} \frac{\pi^2}{3} \frac{1}{Z} \frac{\php}{\ph} 
\end{equation}
Hence, odd-frequency scattering does not contribute at low temperature. 
The conventional elastic value Eq.~(\ref{eq:elastic}) is recovered, with the notable difference that the prefactor 
is enhanced by the effective mass $1/Z=m^*/m$. 
Indeed, it was emphasized in Ref.~\cite{behnia_2004} that $S/T$ is proportional to the 
linear term in the specific heat ($\sim m^*/m$) in many materials. 
%

The crossover between the low-$T$ elastic limit and the high-$T$ inelastic limit 
is illustrated on Fig.~\ref{fig:fl}. The data in this plot as well as in Fig.~\ref{fig:spl} and Fig.~\ref{fig:pl} are obtained from evaluating the Seebeck coefficient 
with Eq.~(\ref{eq:Seebeck}). 
In appendix~\ref{app:kubo}, we show that the results are unchanged when evaluated from 
the full Kubo formula.
We see that, in a Fermi liquid, the band value of the Seebeck coefficient enhanced by the effective mass effect ($Z$) 
is recovered below the crossover temperature $T^*$. In that regime, the particle-hole asymmetry of the inelastic scattering 
has no influence on the Seebeck coefficient. 
In practice, $T^*$ can be estimated as the characteristic temperature at which the measured $T$-dependent contribution to the resistivity becomes comparable in magnitude to the residual resistivity at low-$T$.

\subsection{ `Skewed' Non-Fermi liquids.} 
The washing out of the effects of the particle-hole asymmetry in the Fermi liquid case happens because the odd-frequency terms in the inelastic scattering rate are 
subdominant in comparison to the even ones, and hence do not contribute to $S$ at low-$T$. 
We consider now a class of non-Fermi liquids in which, in contrast, the odd-frequency terms 
are of the same order as the even-frequency ones, such 
that the inelastic scattering rate obeys a scaling form:
\begin{equation}
    \gamin(T,\omega) = \lambda (\pi T)^\nu \, g\left(\frac{\omega}{T}\right).
    \label{eq:scaling}
\end{equation}
Here, $\nu$ is an exponent (we focus on $\nu\leq 1$ in the following) and 
the scaling function $g(x)$ contains both an even and an odd component. 
It has a regular expansion at small $x$, so that for $\omega\lesssim T$: 
$\gamin\sim \lambda (\pi T)^\nu g(0) + g^{\prime}(0) T^{\nu-1}\omega + \cdots$, 
while $g(x)\sim c_{\pm}^\infty\cdot|x|^\nu$ at large $x$. 
The effects discussed in this article do not depend on the specific form of the scaling function $g(x)$. 
Systems obeying this scaling form with a non-even scaling function can be called `skewed non-Fermi liquids'. 

It is of course expected that systems such as doped Mott insulators 
should display a particle-hole asymmetry (see e.g.~\cite{Deng_2013}). However, 
while it is clear that such an asymmetry exists on energy scales comparable to 
electronic ones, it is a more demanding requirement that this asymmetry 
persists down to frequencies comparable to temperature itself, 
as assumed in (\ref{eq:scaling}). Indeed, this does not apply in a Fermi liquid.  
In contrast, we note that the scaling form Eq.~(\ref{eq:scaling}) has been shown to apply in 
overscreened Kondo models controlled by a non-Fermi liquid fixed 
point~\cite{parcollet_kondo_prb_1998}. 
It is also relevant to the proximity of the quantum critical point of doped 
random-exchange Hubbard models, related to Sachdev-Ye-Kitaev (SYK) models~\cite{SY,kitaev_talk}, 
as recently studied in Refs.~\cite{georges_qsg_prb_2001,parcollet_doped_1999,
Sachdev_2015,davison_prb_2017,Tikhanovskaya_I_2020,Tikhanovskaya_II_2020,
Kruchkov_2020,Dumitrescu_2021}. 

For such models obeying conformal invariance at low-energy, $g(x)$ is a universal scaling function that depends only on the 
exponent $\nu$ and on a `spectral asymmetry' parameter $\alpha$. 
Its exact form 
was derived in Ref.~\cite{parcollet_kondo_prb_1998} and reads 
(with the normalisation $g_{\alpha=0}(0)=1$): 
\begin{equation}
  g(x)\,=\,\bigg|\Gamma \left[\frac{1+\nu}{2}+ i \frac{x+\alpha}{2\pi}\right]\bigg|^2
    \frac{\cosh(x/2)}{\cosh(\alpha/2)\Gamma\left[(1+\nu)/2\right]^2}, 
    \label{eq:cft_scaling}
\end{equation}
where $\Gamma(x)$ is the $\Gamma-$function. 
%
The $\omega/T$ scaling form is also relevant to the proximity of a quantum critical point 
associated with a strong-coupling fixed point~\cite{sachdev_book}.
\begin{figure}
 \begin{center}
   \includegraphics[width=1.0\columnwidth,keepaspectratio]{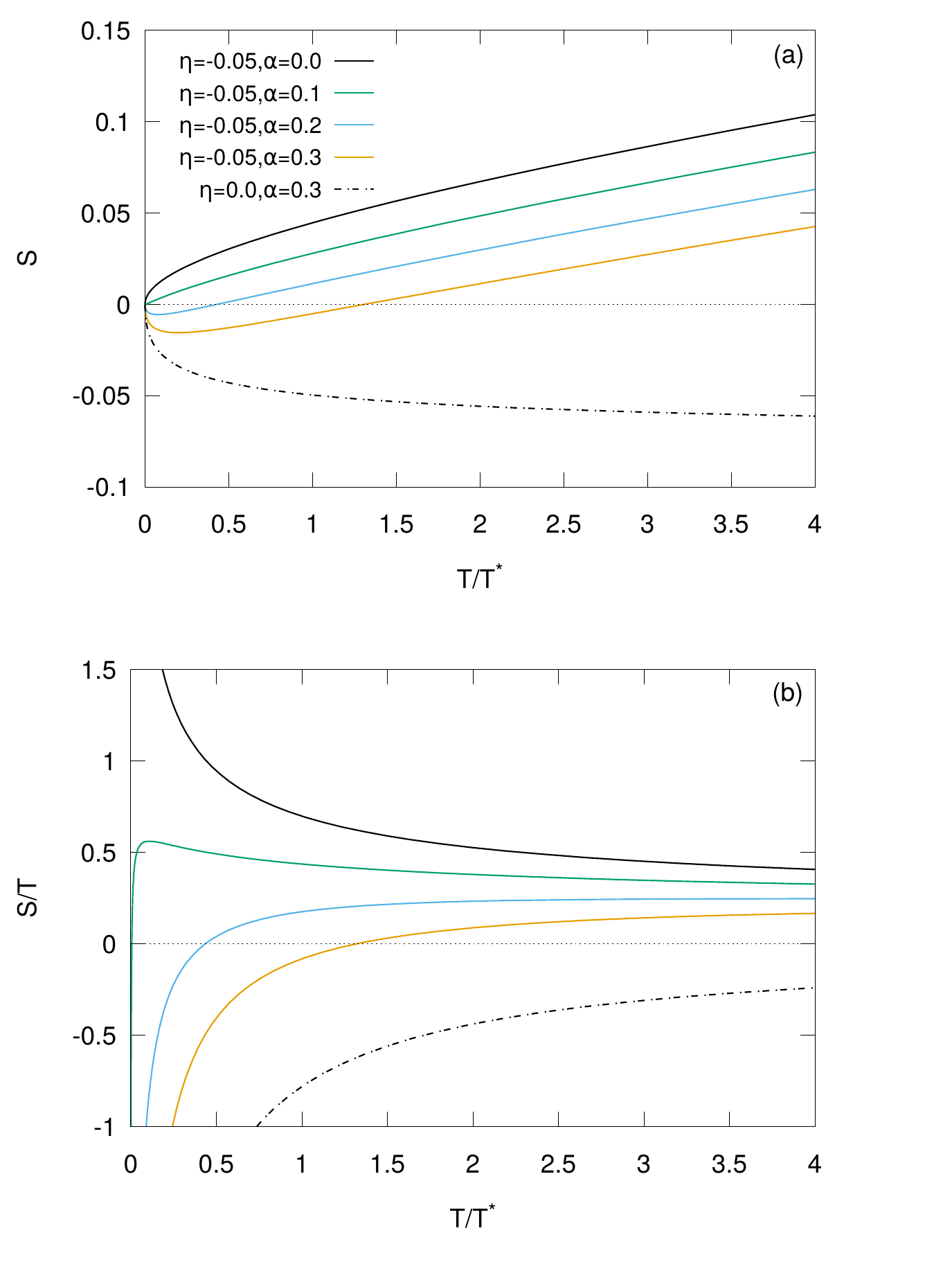}
   \end{center}
   \caption{\label{fig:spl} 
   `Skewed' non-Fermi liquid with $\nu=1/2$. Panel (a) shows the Seebeck coefficient vs. $T/T^*$ in units of $k_B/e$, for $\pi T^*/\gamma= 0.2$ and several values of 
   the asymmetry parameter $\alpha$. Panel (b) shows $S/T$ in units of $k_B/(e \gamma)$.  
   At low-$T$, the sign of $S$ is seen to depend on $\alpha$, while 
   at high-$T$ all curves have the same sign, set by the bandstructure $\eta=\gamma\php/\ph>0$ (electron-like). 
   The dashed-dotted curve corresponds to the special case $\eta=0$ in which the sign of $S$ is determined 
   solely by the scattering rate asymmetry (see text).
   }
 \end{figure}

The crossover temperature $T^*$ separating the `elastic' and `inelastic' 
limits now reads $(\pi T^*)^\nu = \gamel/\lambda$. 
In the low-$T$ elastic limit, $\taup\sim 1/\gamel$ and $\taum\sim - \lambda\gamma^{-2} (\pi T)^\nu g_{-}(x)\sim -\gamma^{-1} (T/T^*)^\nu g_{-}(x)$. 
In this expression, $\gm(x)=[g(x)-g(-x)]/2$ is the odd-frequency 
component of the scaling function $g$.  
For the Seebeck coefficient at low $T$ one  obtains
\begin{equation}
    S\big|_{\el}^{NFL}
     =-\frac{k_B}{e} \left[ \frac{\pi^2 }{3}\frac{\php}{Z(T) \ph } T  - \frac{\lambda (\pi T)^\nu}{\gamma} \cm \right], 
\end{equation}
in which $\cm=\braket{x\gm(x)}$ is a universal constant depending only on $g$. Ignoring at first the temperature dependence of quasiparticle weight $Z(T)$ one sees that whenever $\nu \leq 1$,  the particle-hole asymmetry of the inelastic scattering influences $S\big|_{\el}^{NFL}$ even at the lowest temperatures, 
in sharp contrast to the Fermi liquid case discussed above.

What is the influence of the temperature dependence of $Z$?
For $\nu<1$ (we consider separately the $\nu=1$ case below), $Z$ vanishes at low-$T$  
as $\sim T^{1-\nu}$, and it can be shown (see Appendix~\ref{app:skewed}) 
that $T/Z(T)\sim \lambda (\pi T)^\nu/\prefZ + T$ with $\prefZ$ 
a universal constant depending only on the scaling function $g$. 
Hence, we see that, remarkably, 
the two terms in the numerator $I_1$ of Eq.~(\ref{eq:Seebeck}) have the same $T$-dependence: 
$T\braket{x^2\taup}/Z\sim T^\nu$ and $\braket{x\taum}\sim T^\nu$. One therefore obtains 
for $T\ll T^*$: 
%
%
\begin{eqnarray}
    S\big|_{\el}^{NFL}
    &=&-\frac{k_B}{e} \frac{\lambda}{\gamma} (\pi T)^\nu\,\left[\frac{\pi^2}{3\prefZ}\gamma\frac{\php}{\ph}-\cm\right] \nonumber \\
    &=&-\frac{k_B}{e} \left(\frac{T}{T^*}\right)^\nu\,\left[\frac{\pi^2}{3\prefZ}\eta-\cm\right]
\label{eq:Seebeck_lowT_NFL}
\end{eqnarray}
The dimensionless parameter $\eta\equiv \gamma\php/\ph$ is the ratio of the elastic scattering rate to the 
characteristic energy scale associated with the bandstructure asymmetry. 
Hence, in this case, $S$ behaves as $T^\nu$ at low-$T$, corresponding to a divergent slope $S/T$. 

Expression Eq.~(\ref{eq:Seebeck_lowT_NFL}) has several remarkable features. Firstly, we see that the odd-frequency inelastic scattering 
contributes to the low-$T$ Seebeck on equal footing  with the even-frequency/elastic contribution. 
Both terms in Eq.~(\ref{eq:Seebeck_lowT_NFL}) have the same $T$-dependence $\sim T^\nu$ but for 
different reasons: the first one because of the vanishing of $Z(T)$, and the second one because of the 
$T$-dependence of $\gamin$. 
Secondly, in contrast to the former, this odd-frequency contribution is 
completely independent of the band-structure asymmetry: its sign is dictated by that of the constant $\cm$, and 
thus by the intrinsic asymmetry of the inelastic rate scaling function. 
If this term dominates, the overall sign of the Seebeck can be opposite to that predicted by band-structure considerations 
in the low-$T$ limit, even when elastic scattering is present. 
This is one of the main results of this work. 
The odd-frequency contribution dominates over the first term when the dimensionless ratio $\eta=\gamma\php/\ph$ is small, i.e. for clean-enough systems. 
Thirdly, we note that $S$ depends on both the inelastic constant $\lambda$ and the elastic rate $\gamma$. 
This is an unusual situation in which the strength of the scattering does not drop out of the value of $S$ at low-$T$. 

%
The behavior of Eq.~\ref{eq:Seebeck} with temperature is investigated analytically 
in Appendix~\ref{app:T}. At high $T$, we obtain
\begin{equation}
S \sim -\frac{k_B}{e}\left[ 
    \eta\frac{c_2}{c_0}\frac{T}{\gamel} + 
    \eta\frac{c_2}{c_0\prefZ}\left(\frac{T}{T^*}\right)^\nu\,+\,
    \frac{c_1}{c_0}\right]\label{eq:ht1}
\end{equation}
with $c_n=\braket{x^n/g}$. 
Hence, as long as $\eta=\gamma\php/\ph\neq 0$, the behavior of $S$ at high temperature 
is dominated by the first term of Eq.~(\ref{eq:ht1}) (with a linear dependence in $T$, the same sign as the band-structure result, but with a slope that is corrected by $c_2/c_0$). In contrast to the low-$T$ limit, the sign of the Seebeck coefficient is thus given by the band-structure term.
Remarkably, in the absence of elastic scattering or band asymmetry ($\eta=0$), $S$ instead tends to a constant value 
$- c_1/c_0\, k_B/e$ which depends only on the universal scaling function and its asymmetry. 
A related finding was reported in Ref.~\cite{davison_prb_2017} (see also \cite{Kruchkov_2020}) 
in the context of SYK models, where $S$ was shown to be constant 
and determined by the spectral asymmetry $\alpha$ (related by holography to the electric 
field/charge of the black hole and to the ground-state entropy of these models). 

On Fig.~\ref{fig:spl}(a,b), we display the temperature dependence of $S$ and $S/T$, respectively,  
plotted vs.~$T/T^*$ in the non-Fermi liquid case for several values of the asymmetry parameter $\alpha$ 
in Eq.~(\ref{eq:cft_scaling}). 
We see that at low-$T$ the sign of $S$ can be changed by the scattering rate asymmetry, while it is set 
by the bandstructure term at high-$T$, except for vanishing $\eta$ where the Seebeck approaches a constant given by the last term of Eq.~\eqref{eq:ht1}, instead. 

The overall behavior at larger $T$ reveals a $\mathrm{const}+ T^\nu + T$ 
behavior at higher $T$ (with $\nu=1/2$ in Fig.~\ref{fig:spl}), with a different coefficient of the $T^\nu$ term in the high-$T$ regime from that in the low-$T$ regime. 
We see that the larger the asymmetry, the larger the temperature at which the sign change 
occurs as compared to the symmetric case.

A remark is in order here. In microscopic models or materials realizations, the universal scaling form of the scattering rate (\ref{eq:scaling}) is only expected to apply  
below a certain cutoff, which is of order of the bare electronic energy scales, e.g. the bandwidth. 
The physics discussed here obviously can only apply for temperatures below that cutoff. 
Hence, for our analysis of the $T$-dependence to be valid, the system should be clean enough that 
the crossover scale $T^*$ is smaller than such bare electronic energies at which 
non-universal effects will appear.

\subsection{Skewed Planckian metal.}
We now discuss the case of a `Planckian' metal~\cite{Zaanen04,Marel2003,Bruin_2013,Hartnoll_2014,Legros19,Grissonnanche_2020,Varma_2020} ($\nu=1$)  
with a particle-hole asymmetry scaling with $\omega/T$, so that  
$\gamin=\pi\lambda g(0)\,T + \pi\lambda g^\prime(0)\,\omega +\cdots$ at low $\omega$ and $T$ 
(`skewed' Planckian metal). 
%
Setting $\nu=1$ in expression Eq.~(\ref{eq:Seebeck_lowT_NFL}) would predict a linear dependence of $S$ at low-$T$, with a slope that is given in terms 
of mutual effect of the band-structure and inelastic stattering rate asymmetries. This behavior corresponds to a case where quasiparticle weight is approximated by a constant.
However, 
the quasiparticle weight actually vanishes logarithmically: 
$1/Z(T)\sim 1+\pi\lambda \prefZ^{-1} \ln \Lambda/T$, with $\Lambda$ a high-energy cutoff and $\prefZ$ 
a universal constant depending on the scaling function $g$, see Appendix~\ref{app:skewed}      
(note that $\lambda$ is dimensionless in the Planckian case). 
Hence, in the low-$T$ `elastic' limit $T\ll T^*= \gamel/\pi\lambda$:
\begin{equation}
\frac{S}{T}\bigg|_{\el}^{SPM} \sim -\frac{k_B}{e}\,\left[
\pi\lambda\frac{\pi^2}{3c_Z}\frac{\php}{\ph}\ln\frac{\Lambda}{T}+
\frac{\pi^2}{3}\frac{\php}{\ph}-\cm\frac{\pi\lambda}{\gamel}
+\cdots 
\right]
\label{eq:Seebeck_slope_lowT_Planck}
\end{equation}
Thus, in a Planckian metal, $S/T$ ultimately diverges logarithmically at very low-$T$. 
This corresponds to the logarithmic divergence of the effective mass (specific heat coefficient). 
The sign of the logarithmic term in $S$ is dictated by bandstructure. 
In contrast, the last term in Eq.~(\ref{eq:Seebeck_slope_lowT_Planck}) is controlled by the odd-frequency part of the inelastic scattering, 
and has a sign which can counteract the conventional bandstructure effect corresponding to the second term in Eq.~(\ref{eq:Seebeck_slope_lowT_Planck}). 
Interestingly, we note that the `skewed' term dominates over the conventional one for cleaner systems 
$\eta/\lambda\lesssim 3c_{-}/\pi$. 
The full temperature dependence is again given by Eq.~(\ref{eq:crossover}), setting $\nu=1$ and replacing $\prefZ$ 
by $\prefZ/\ln\Lambda/T$. 

\begin{figure}
 \begin{center}
   \includegraphics[width=1.0\columnwidth,keepaspectratio]{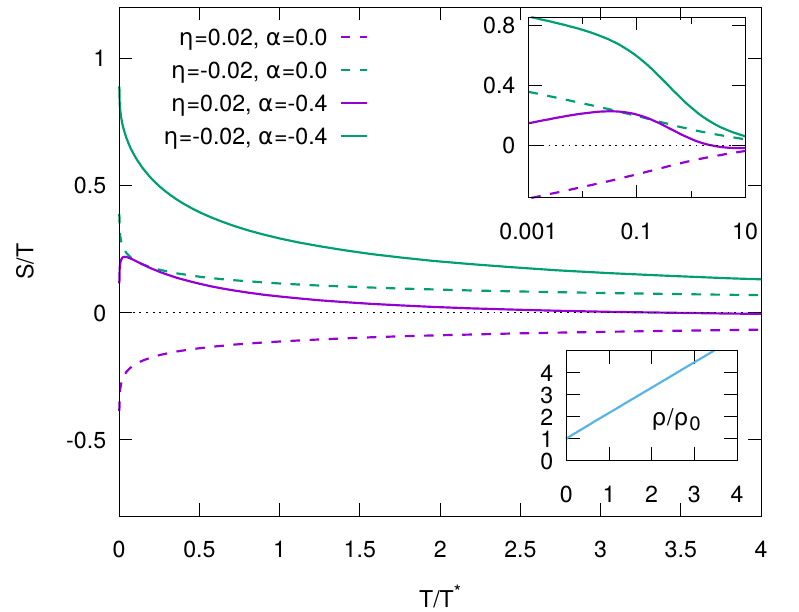}
   \end{center}
   \caption{\label{fig:pl} Planckian metal. 
   Temperature dependence of $S/T$ (in units of $k_B/(e \gamel)$) for two opposite values of the bandstructure term 
   $\eta=\gamel\php/\ph$: electron-like $\eta>0$ and hole-like $\eta<0$.  
   Plain (resp. dahed) lines are for a particle-hole asymmetric (resp. symmetric) scattering rate 
   ($\alpha\neq 0$, resp. $\alpha=0$). $\pi T^*/\gamel=1$.
   Top inset: $S/T$ on a logarithmic scale, emphasizing the low-$T$ behaviour.
   Bottom inset: Linear dependence of the resistivity $\rho/\rho(T=0)=I_0(0)/I_0(T)$ on $T/T^*$. 
   } 
 \end{figure}

To illustrate these effects, we display on Fig~\ref{fig:pl} the $T$-dependence of $S/T$ for a `skewed' Planckian metal.  
Two opposite signs of the bandstructure term $\php/\ph$ are considered. In the absence of a scattering rate asymmetry ($\alpha=0$) 
it is seen that $S/T$ has a rather weak $T$-dependence (except at low-$T$ where the logarithmic term becomes relevant), while 
it acquires significant $T$-dependence for $\alpha\neq 0$. Importantly, in the presence of an asymmetry, 
the sign of the Seebeck coefficient can be reversed 
in comparison to its bandstructure value over a wide range of temperature. 


\section{Summary and Discussion}
\label{sec:disc}
 In summary, our work reveals the importance of a particle-hole asymmetry of the inelastic scattering rate in a class of non-Fermi liquids. This asymmetry influences the Seebeck coefficient even in the asymptotic low-temperature regime, 
 when inelastic scattering events are comparatively rare as compared to elastic scattering on impurities.
%
%

A number of materials display non-Fermi liquid behaviour and an unconventional $T$-dependence of the Seebeck coefficient. 
This has been addressed in previous theoretical work, such as  Refs~\cite{Paul_2001} and \cite{Buhmann_2013} 
which have emphasized the logarithmic divergence of $S/T$ in metals with a $T$-linear scattering rate. 
The $T$-dependence of $S$ close to a Fermi surface Lifshitz transition has been considered in \cite{Varlamov_1989,Herman_2019}. 
However, to our knowledge, the key role of a particle-hole asymmetry of the inelastic scattering rate 
in non-Fermi liquid metals with $\omega/T$ scaling has not been discussed before. 
The theory presented here may be relevant when both the temperature dependence and the sign of the Seebeck coefficient 
are found to be unconventional. 

Cuprate superconductors in the `strange metal' regime display clear signatures of quantum criticality: 
T-linear resistivity (for reviews see e.g.\cite{Hussey_2008,Taillefer_2010,Varma_2020}), 
a logarithmic divergence of the specific heat coefficient $C/T$~\cite{Michon_2019}, 
and $\omega/T$ scaling observed in optics~\cite{Homes_2004} and angular-resolved photoemission 
(ARPES)~\cite{Reber_2019} spectroscopies. 
An increase of the in-plane $S/T$ at low-$T$ reminiscent of Fig.~\ref{fig:pl} has been reported 
for La$_{1.8-x}$Eu$_{0.2}$Sr$_x$CuO$_4$ (Eu-LSCO)~\cite{Laliberte_2011} at hole doping $p\simeq 0.21$ and $p\simeq 0.24$ 
and for La$_{1.6-x}$Nd$_{0.4}$Sr$_x$CuO$_4$ (Nd-LSCO)~\cite{Daou_2009,Collignon_2020} at $p\simeq 0.24$, just above 
the critical doping $p^*$ at which the pseudogap phase terminates. 
Interestingly, $S$ was found to be positive at those doping levels, while simple considerations 
based on band-structure and isotropic elastic scattering 
yield a negative value~\cite{Verret_2017}. 
It is thus tempting to infer from these observations that the quantum critical (strange metal) 
regime of those cuprate superconductors may be described as a `skewed Planckian metal'. 
We emphasize that experiments involving different controlled levels of disorder would 
play a decisive role in assessing the relevance of the mechanism proposed here, since the 
asymmetric term in Eq.~(\ref{eq:Seebeck_slope_lowT_Planck}) becomes larger for cleaner systems. 
For a discussion of the relevance of the present theory to the interpretation of recent measurements of the ab-plane 
and c-axis measurements of the Seebeck coefficient in Nd-LSCO, see Ref.~\cite{gourgout_2021}.

In Eu-LSCO, Nd-LSCO and also Bi-2201~\cite{Lizaire_2020}, the increase of $S/T$ at low-$T$ observed 
experimentally appears consistent with a logarithmic dependence. 
Given the sign of the bandstructure term for Nd-LSCO, one would expect from 
Eq.~(\ref{eq:Seebeck_slope_lowT_Planck}) a negative coefficient of this logarithmic term, in contrast 
to the experimental observation. This may 
suggest that the asymmetry term is actually 
dominant (as on  Fig.~\ref{fig:pl}) in the range of temperature of the measurement, 
in which the increase of $C/T$ is moderate~\cite{Michon_2019}. 

As emphasized recently by Jin and coworkers~\cite{jin_behnia_2021}, in the overdoped regime the Seebeck coefficient of the LSCO family remains 
positive~\cite{Nakamura_1993,Zhou_1995}. 
These overdoped compounds behave as Fermi liquids at low enough temperature however~\cite{Nakamae_2003,Fatuzzo_2014,Horio_2018}. 
Hence, in view of the discussion above (Fig.~\ref{fig:fl}), 
a particle-hole asymmetry of the {\it inelastic} scattering rate is unlikely to be responsible 
for this unexpected sign of $S$. 
It is possible 
that a particle-hole asymmetric dependence of the {\it elastic} scattering rate on momentum and energy is relevant to explain this observation~\cite{Hussey_2008,Narduzzo_2008,Fang_2020}. 
We also note that the Seebeck coefficient of other single-layer cuprates such as Hg1201~\cite{Yamamoto_2001} 
and Bi2201~\cite{Konstantinovic_2002,Kondo_2005} has been reported to be negative in the overdoped regime.

Recently, the present theory was applied to experimental 
measurements of the ab-plane and c-axis 
Seebeck coefficients in Nd-LSCO~\cite{gourgout_2021}. 
The scattering rates used in this analysis were extracted from angular-dependent magneto-resistance measurements, which reveal that 
the elastic (temperature-independent) rate is momentum dependent 
but that the inelastic one is not~\cite{Grissonnanche_2020}. 
The theory was hence minimally extended as compared to the present work by using a  momentum-dependent elastic lifetime and found to describe the experimental results well. We stress that the qualitative aspects (change of sign for the in-plane Seebeck coefficient due to inelastic scattering) are already well described by the simpler approach used here  which neglects the momentum dependence, but that quantitative aspects 
require the momentum dependence to be taken into account, 
especially for the out-of-plane Seebeck coefficient .

Our results are relevant also in the context of the non-Fermi liquid 
quantum critical point separating a metallic spin-glass phase 
and a Fermi liquid metal in doped SYK models, which has attracted a lot of attention 
recently~\cite{Joshi_2020,Cha_2020,Shackleton_2020,Tikhanovskaya_I_2020,Tikhanovskaya_II_2020,
Dumitrescu_2021}. 
Indeed, numerical results show strong spectral asymmetry in the quantum critical regime, 
possibly signalling a skewed non-Fermi liquid at the critical point~\cite{Dumitrescu_2021}. 
We also note that in the context of transport in SYK models, the Seebeck coefficient 
has been emphasized as a probe of the universal ground-state entropy~\cite{davison_prb_2017,Kruchkov_2020}. 

Our work may also have relevance to twisted bilayer graphene and related systems, 
in which linear resistivity and Planckian behaviour have been observed, see e.g.~\cite{Park_Diffusivity_2020}. 
A systematic investigation of thermoeletric effects and heat transport in these materials would be of great interest. 
We note that a recent study of twisted bilayer graphene reports a rich temperature dependence with changes of sign in the Planckian regime~\cite{Ghawri_2020}.

\acknowledgements

We are grateful to Amir Ataei, Nicolas Doiron-Leyraud, Adrien Gourgout, 
Ga\"{e}l Grissonnanche, Louis Taillefer and Simon Verret 
for sharing their experimental data and for a collaboration and discussions on this topic. 
We also acknowledge discussions with Kamran Behnia, Christophe Berthod, Philipp Dumitrescu, Pablo Jarillo-Herrero, Olivier Parcollet, Subir Sachdev, 
Nils Wentzell and Manuel Zingl. 
A.G. acknowledges support from the European Research Council (ERC-QMAC-319286).  
J.M. acknowledges funding by the Slovenian Research Agency (ARRS) under Program No. P1-0044, J1-1696, and J1-2458. 
The Flatiron Institute is a division of the Simons Foundation.  

%
\appendix

\section{Kubo formula: Thermopower and Conductivity}
\label{app:kubo}
The  conductivity $\sigma$, thermopower $S$ and (electronic) heat conductivity $\kappa$ are given by: 
\begin{equation}
\sigma = e^2 L_{11}\,\,\,,\,\,\,
S=-\frac{L_{12}}{e L_{11}}\,\,\,,\,\,\,
\kappa = T\left[L_{22}-\frac{L_{12}^2}{L_{11}}\right]
\end{equation}
in which $L$ are Onsager's coefficients and $e$ is the absolute magnitude of  the electron charge. 
In the following, we set for simplicity $\hbar=k_B=e=1$ (except when restored in final results).  
Within the Kubo formalism, and neglecting vertex corrections, the Onsager coefficients are given by (denoting for 
simplicity $L_{11}\equiv L_0, L_{12}\equiv L_1, L_{22}\equiv L_2$: 
\begin{equation}
L_n\,=\frac{1}{T^n} \int d\omega \left(-\frac{\partial f}{\partial\omega}\right)\,\omega^n\, {\cal T}(\omega)
\end{equation}
in which $f(\omega)=1/[1+e^{\omega/T}]$ is the Fermi function and 
\begin{equation}
{\cal T}(\omega) = 2 \pi \int \frac{d^d k}{(2\pi)^d}\, v_{\vk}^2\,A(\vk,\omega)^2
\end{equation}
In this expression, $v_{\vk}=\left(\nabla_{\vk}\ek\right)_\alpha/\hbar$ 
denotes the band velocity in the direction $\alpha=x,y,z$ being considered, and 
$A(\vk,\omega)$ is the electronic spectral function, related to the self-energy 
$\Sigma=\Sigma(\omega+i0^+,\vk)$ by:
\begin{equation}
A(\vk,\omega) = \frac{1}{\pi} \frac{\Gamma/2}{\left(\omega+\mu-\ek-\res\right)^2+(\Gamma/2)^2}
\end{equation}
In this expression, $\Sigma=\Sigma(\omega+i0^+,\vk)$ is the self-energy due to inelastic interactions between electrons 
and $\Gamma$ is the full scattering rate including both elastic and inelastic terms:
\begin{equation}
\Gamma(\vk,\omega)\,=\,\gamma_{\vk} - 2\ims(\vk,\omega+i0^+)
\end{equation}
These expressions can be further simplified when the elastic scattering rate $\gamma_{\vk}$ and electron-electron
self-energy $\Sigma$ do not depend on momentum and when the  
 vertex corrections can be neglected~\cite{Khurana_1990}.

Using the band transport function $\Phi$ one can express
\begin{equation}
{\cal T}(\omega) = \pi \int d\ee\, \Phi(\ee)\, A(\ee,\omega)^2
\end{equation}
We change the integration  variable by setting $\ee=\omega+\mu-\res+\frac{\Gamma}{2}\,y$, so that:
\begin{equation}
{\cal T}(\omega) = \pi \frac{2}{\Gamma}\, \int_{-\infty}^{+\infty} dy\, \Phi\left(\omega+\mu-\res+\frac{\Gamma}{2}\,y \right)\, 
\left(\frac{1/\pi}{y^2+1}\right)^2
\end{equation}
We now perform an expansion of this expression for small $\Gamma$ and retain only the most singular term, yielding:
\begin{equation}
{\cal T}(\omega) = \frac{1}{\Gamma}\, \Phi\left(\omega+\mu-\res\right)+\cdots
\end{equation}
where we have used $\int_{-\infty}^{+\infty} dy\, 1/\pi^2(y^2+1)^2=1/(2 \pi)$. 
It is convenient to introduce the notations:
\begin{eqnarray}
\overline{\ee}_F(T)\equiv \mu -\res(\omega=0,T) \nonumber \\ 
1-\frac{1}{Z(T,\omega)}\equiv \frac{1}{\omega}\left[\res(\omega,T)-\res(0,T)\right]
\end{eqnarray}
so that, to dominant order in $\Gamma$:
\begin{equation}
{\cal T}(\omega) = \frac{1}{\Gamma(T,\omega)}\, \Phi\left[\overline{\ee}_F(T)+\frac{\omega}{Z(T,\omega)}
\right]+\cdots
\end{equation}
Inserting this expression in the Onsager coefficients $L_n$ above, and changing variable to $x=\omega/T$ in the integral 
over frequency, we obtain, with $\tau=1/\Gamma$:
\begin{equation}
L_n\,= \int dx\,\frac{x^n}{4\cosh^2\frac{x}{2}}\,
\Phi\left[\overline{\ee}_F(T)+x\,\frac{T}{Z(T,xT)}\right]\,\tau(T,xT)
\end{equation}
We now perform a low-$T$ expansion of this expression. We note that for both the Fermi liquid and non-Fermi liquid cases 
considered in this article, $T/Z(T,xT)$ vanishes at low-$T$ (as $\sim T$ and $\sim T^\nu$, respectively). We assume 
furthermore that the $T$-dependence of $\overline{\ee}_F(T)$ is subdominant and can be neglected. We thus obtain: 
\begin{equation}
L_n\,=\,\Phi_0\,\braket{x^n\tau(T,xT)}\,+\,T\,\Phi^\prime_0\,\braket{x^{n+1}\frac{\tau(T,xT)}{Z(T,xT)}}+\cdots
\end{equation}
with the notation:
\begin{equation}
\langle F(x) \rangle\,=\,\int_{-\infty}^{+\infty}dx\, \frac{F(x)}{4\cosh^2\frac{x}{2}}
\end{equation}
Retaining just the lowest two terms in the Taylor expansion of $\phi(\varepsilon)$ is sufficient at low temperatures as the transport function usually varies only on the bare electronic scales, e.g. the bandwidth.
We have checked furthermore that neglecting the frequency ($x$) dependence of $Z$ in these equations is a good 
approximation, so that we finally obtain the expressions Eqs.~(\ref{eq:Seebeck},\ref{eq:I1},\ref{eq:I0}) of Sec.\ref{sec:meth}.

%

We also note the expression of the conductivity: 
\beq
\sigma = e^2\ph\,I_0(T)\,\simeq\,e^2\ph\,\braket{\taup}
\eeq
Although the non-Fermi liquid case does not have conventional quasiparticles,  the final expressions for $S$, $\sigma$ and the Onsager coefficient $L_n$'s at low-$T$ are 
formally identical to the ones obtained by applying a Boltzmann equation formalism to excitations 
with a lifetime $\tau^{ex}=\tau/Z$ and 
dispersing as $\ee^{ex}_{\vk}=Z\ee_{\vk}$, corresponding to a transport function of these excitations 
$\Phi_{ex}(\ee)=Z\Phi(\ee/Z)$.

\begin{figure}
 \begin{center}
   \includegraphics[width=1\columnwidth,keepaspectratio]{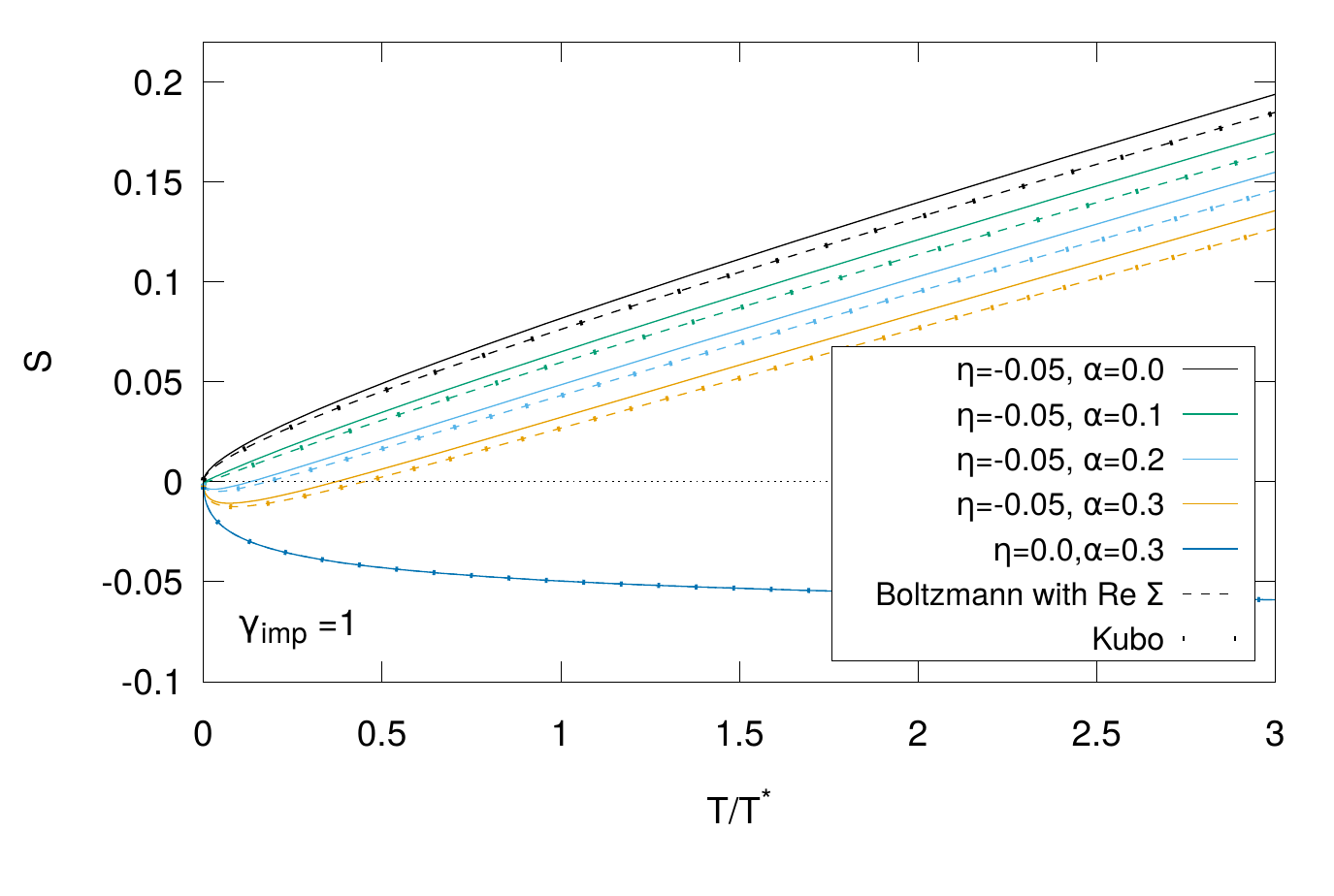}
   \end{center}
   \caption{\label{fig:Kubo_vs_Boltz}
     Seebeck coefficient in units $k_B/e$ for the $\nu=1/2$ case. $\pi T^*/\gamma= 1$.  The results of the Boltzmann calculation with $Z=Z(T)$ (full lines) are compared to the Kubo calculation (dotted) and a Boltzmann calculation that retains the frequency dependence of $Z=Z(\omega,T)$ (dashed). The results of the latter two cannot be distinguished on the scale of this plot. Similar agreement between the Boltzmann and Kubo results is found for all the data in the main text.  
   }
 \end{figure}

On Fig.~\ref{fig:Kubo_vs_Boltz}, we compare the Seebeck coefficient evaluated with the full Kubo formula and with the 
simplified Boltzmann-like expressions. The data is for the subplanckian $\nu=1/2$ case. The results of the full-Kubo evaluation (dotted) agree well with the results of a simplified Boltzmann calculation that approximates $Z(\omega,T)$ by its zero-frequency value $Z(\omega,T) \rightarrow Z(0,T)$ (full lines). The result of such Boltzmann calculations are the data given in the main text. The remaining small discrepancy between the Kubo and the Boltzmann results can be remedied if one retains the frequency dependence of $Z(\omega,T)$ (dashed lines that overlap with the dotted ones).

For application of our theory to cuprates it is essential to retain momentum dependence. However, as recently reported in an angular-dependent magnetoresistance (ADMR) study of cuprates~\cite{Grissonnanche_2020}, the inelastic scattering is found to be local and it is sufficient to retain the momentum dependence in the elastic scattering rate. In such a case, the theory can be minimally extended, with only quantitative but no qualitative changes, by allowing for the elastic rate to have momentum dependence. Such an extension is presented in \cite{gourgout_2021}.

\section{Skewed non-Fermi liquids: $\omega/T$ scaling}
\label{app:skewed}
\subsection{Scaling function and conformally invariant case}
The (local) non-Fermi liquids considered in this article have  a scattering rate that obeys $\omega/T$ scaling ($\nu\leq 1$): 
\begin{equation}
    \gamin(T,\omega) = \lambda (\pi T)^\nu \, g\left(\frac{\omega}{T}\right)
\end{equation}
The scaling function $g(x)$ has a regular expansion at small $x$: $g(x)=g(0)+x\,g^\prime(0)+\cdots$, while at 
large $x$ it obeys $g(x\rightarrow +\infty) \sim c_{+}^\infty x^\nu$ and 
$g(x\rightarrow -\infty) \sim c_{-}^\infty |x|^\nu$. The `skewed' case with a particle-hole asymmetry 
will in general have $g^\prime(0)\neq 0$ and $c_{+}^\infty\neq c_{-}^\infty$. Hence:
\begin{eqnarray}
\gamin(T\gg|\omega|)&\sim&\,\lambda g(0) \,(\pi T)^\nu  + \lambda g^\prime(0) \frac{\pi\omega}{(\pi T)^{1-\nu}} +\cdots \nonumber \\
\gamin(T\ll|\omega|)&\sim& c_{\pm}^\infty\lambda |\pi\omega|^\nu+\cdots 
\end{eqnarray}
Our qualitative results do not depend on the specific form of the scaling function $g(x)$ provided 
it obeys these general properties. 

\begin{figure}
 \begin{center}
   \includegraphics[width=0.49\columnwidth,keepaspectratio]{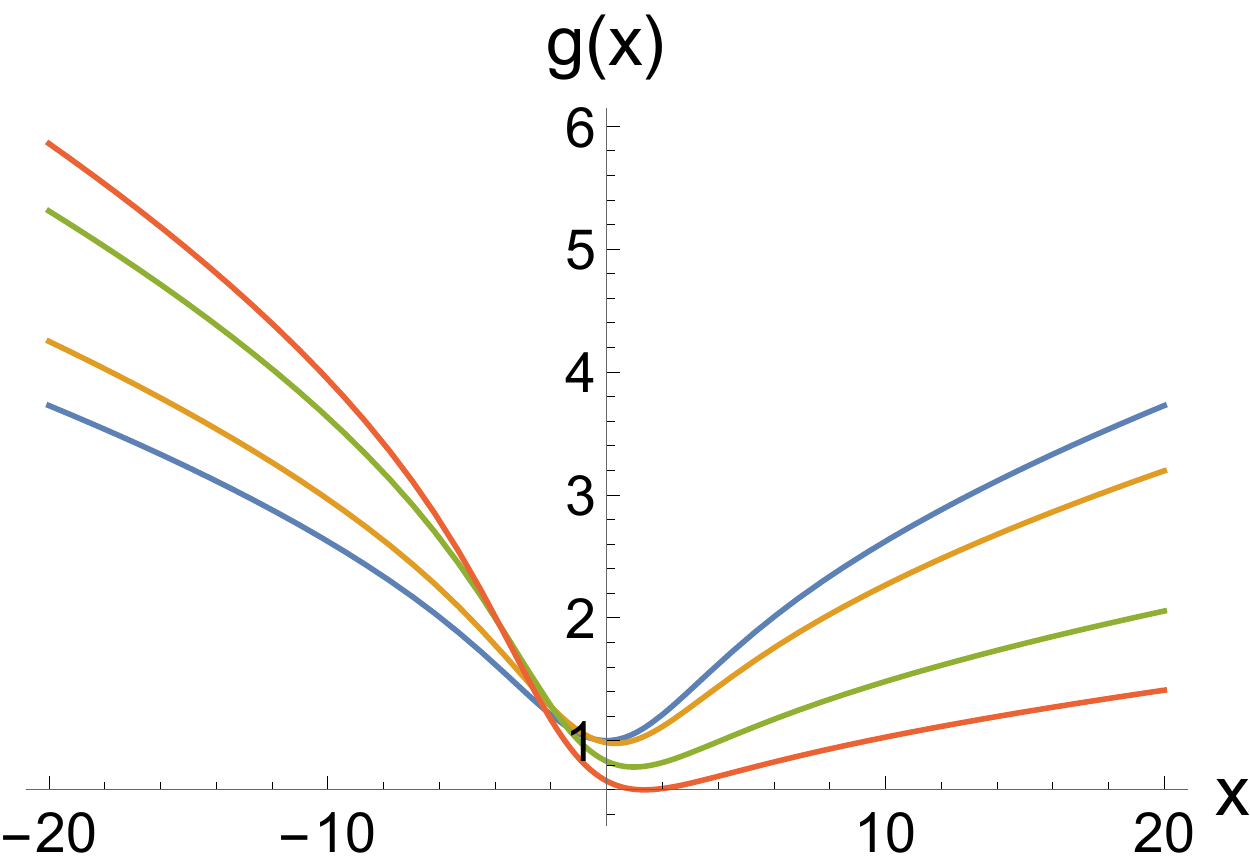}
   \includegraphics[width=0.49\columnwidth,keepaspectratio]{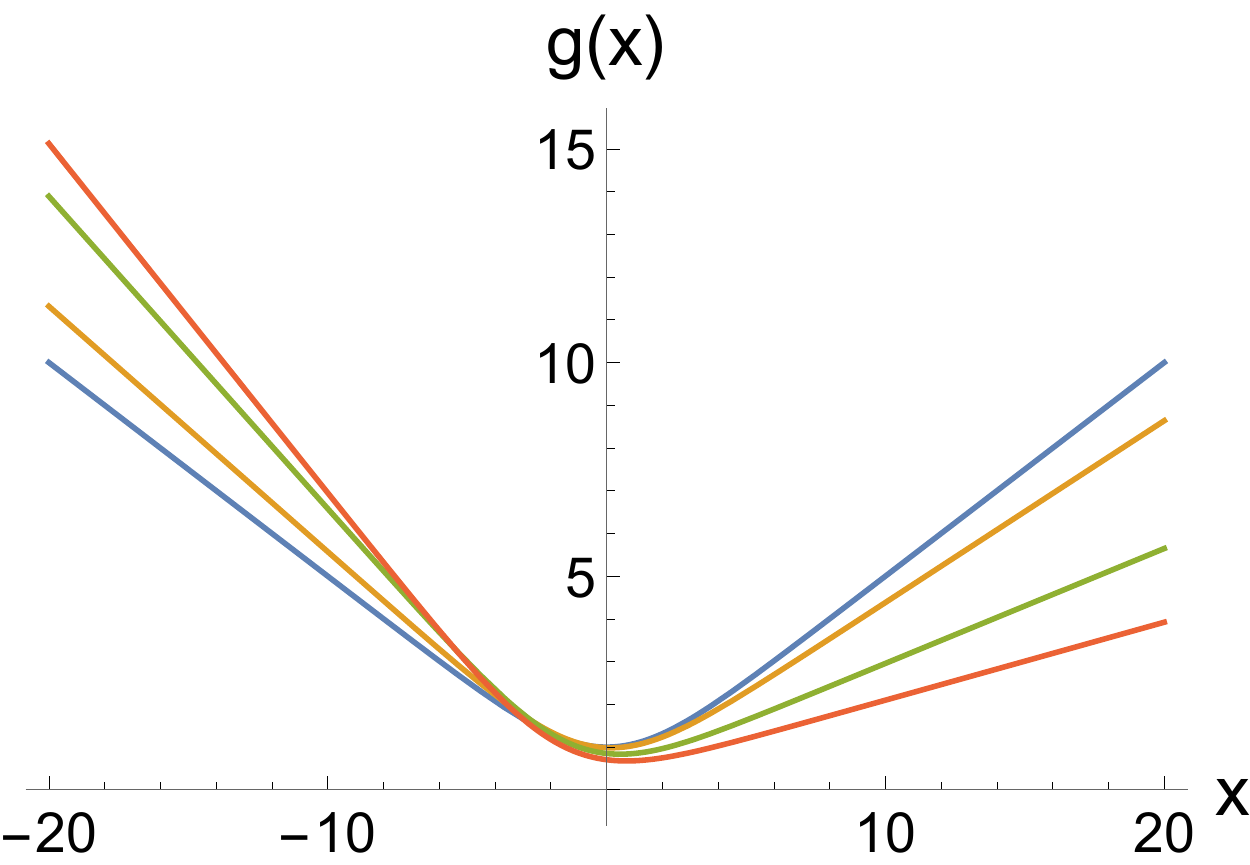}
   \end{center}
   \caption{\label{fig:scaling_function}
Scaling function $g_{\alpha,\nu}(x)$ for $\nu=1/2$ (left) and $\nu=1$ (right), 
for different values of the asymmetry parameter 
$\alpha=0,0.3,1,1.5$ (blue,yellow,green,red).   
 }
 \end{figure}

In Sec.~\ref{sec:fl_vs_nfl} we considered $g(x)$ given by Eq.~(\ref{eq:cft_scaling}). 
It has been shown in \cite{parcollet_kondo_prb_1998}, in the context of overscreened Kondo impurity models, 
that this specific form of the scaling function holds in models which have conformal invariance at low energy/temperature. 
It is also relevant in the context of 
SYK models~\cite{georges_qsg_prb_2001,parcollet_doped_1999,Sachdev_2015,davison_prb_2017,Dumitrescu_2021}. 
The  parameter $\alpha$ controls the particle-hole asymmetry (see Fig.~\ref{fig:scaling_function}), with  
$\alpha=0$ corresponding to particle-hole symmetry $g(x)=g(-x)$ and $g_{-\alpha,\nu}(x)=g_{\alpha,\nu}(-x)$. 
The normalisation $g_{\alpha=0,\nu}(0)=1$ was chosen in the above expression. 
In the Planckian case $\nu=1$, Eq.~(\ref{eq:cft_scaling}) can be cast in the explicit form:
\begin{eqnarray}
g_{\alpha,\nu=1}(x)\,&=&\,\frac{(x+\alpha)/2}{\sinh[(x+\alpha)/2]}\,\frac{\cosh(x/2)}{\cosh(\alpha/2)}  \\ 
 \gamin(T,\omega) &=& \lambda \pi T \, \frac{(\omega/T+\alpha)/2}{\sinh[(\omega/T+\alpha)/2]}\,\frac{\cosh(\omega/2T)}{\cosh(\alpha/2)} \nonumber
\end{eqnarray}
 
The connection between this scaling form and conformal invariance in local models is the following. This invariance 
implies that any imaginary time ($\tau$) fermionic correlation function (for example the self-energy) takes the following 
form in the limit where $\tau$ and inverse temperature $\beta=1/T$ are both large compared to microscopic scales, 
but with arbitrary $\tau/\beta$~\cite{parcollet_kondo_prb_1998}:
\begin{equation}
\Sigma(\tau) \propto e^{\alpha(\tau/\beta-1/2)}\,\left(\frac{\pi/\beta}{\sin \pi\tau/\beta} \right)^{1+\nu}
\end{equation}
and this function has the following spectral representation ($\taubar=\tau/\beta$):
\beq
e^{\alpha(\taubar-1/2)}\,\left(\frac{\pi}{\sin \pi\taubar} \right)^{1+\nu}\,=\,
-\,C\,\int_{-\infty}^{+\infty} dx\,\frac{e^{-x\taubar}}{1+e^{-x}}\,g_{\alpha,\nu}(x)
\eeq 
with $C=\cosh(\alpha/2)(2\pi)^\nu\Gamma[(1+\nu)/2]^2/(\pi\Gamma[1+\nu])$.

Other form of the scaling functions $g$ can also be considered. 
In the absence of a particle-hole asymmetry, the following phenomenological scaling function has sometimes 
been considered in the literature (see e.g \cite{Reber_2019}): 
\begin{equation}
g_{\alpha=0}(x)\,=\,\left[1+x^2/\pi^2\right]^{\nu/2}
\end{equation}
Note that $\nu=2$ reduces to the Fermi liquid scaling function $1+x^2/\pi^2$ - this is also the case of 
Eq.~(\ref{eq:cft_scaling}) for $\nu=2,\alpha=0$. Starting with this form, one can introduce an asymmetry 
for example by deforming the scaling function in an analogous way to Eq.~(\ref{eq:cft_scaling}) \cite{parcollet_kondo_prb_1998}:
\begin{equation}
g_{\alpha}(x)\,=\,g_{0}(x+\alpha)\,\frac{\cosh(x/2)}{\cosh[(x+\alpha)/2]\cosh(\alpha/2)}
\end{equation}
A simple calculation using the spectral representation of $\Sigma$  then shows that:
\begin{equation}
\Sigma_{\alpha}(\tau)\,=\,\frac{e^{\alpha(\tau/\beta-1/2)}}{\cosh\alpha/2}\,\Sigma_{\alpha=0}(\tau)
\end{equation}

\subsection{Scaling form of $Z(T,\omega)$}

Here, we discuss the scaling form of the real part of the self-energy. We start from the spectral representation:  
\begin{eqnarray}
\res(\omega) = P\int d\epsilon \frac{\sigma(\epsilon)}{\omega-\epsilon} \nonumber \\
\sigma(\omega)
= -\frac{1}{\pi}\ims(\omega+i0^+)=\frac{1}{2\pi}\gamin(T,\omega)
\label{eq:resigma_KK}
\end{eqnarray}
in which 'P' denotes the principal part of the integral and we have recalled the relation 
(factor of two) between the imaginary part of the self-energy and the inelastic scattering rate 
for a local theory. The frequency and temperature dependent effective mass enhancement $1/Z(T,\omega)$ is given by:
\begin{equation}
1-\frac{1}{Z(T,\omega)}\equiv \frac{1}{\omega}\left[\res(\omega)-\res(0)\right] 
= P\int d\epsilon \frac{\sigma(\epsilon)}{\epsilon(\omega-\epsilon)}
\end{equation}
Because we have subtracted $\res(0)$, we can substitute the scaling form of $\sigma=\gamin/2$ in the integral without encountering divergencies for a NFL with $\nu<1$,  
and thus obtain:
\begin{equation}
\frac{T}{Z(T,\omega=xT)}\,
=\,T-\frac{\lambda}{2\pi} (\pi T)^\nu \,P\int dy \frac{g(y)}{y(x-y)}
\end{equation}
Hence, remarkably, $T/Z$ obeys a universal scaling form which depends only on the scaling 
function $g$. This is in contrast to a Fermi liquid in which $Z$ depends on all energy scales (as signalled by the fact that inserting the low-energy expression 
$\sigma(\omega)\propto \omega^2+(\pi T)^2$ in Eq.~(\ref{eq:resigma_KK}) would 
lead to a divergent integral). 

For results given in Sec.~\ref{sec:fl_vs_nfl}, we ignore the $\omega$ (i.e. $x$) dependence of $Z$ and replace $Z(T,\omega)$ by $Z(T,0)=Z(T)$. Comparison with the full Kubo formula calculations given in Fig.~\ref{fig:Kubo_vs_Boltz} justify this approximation for transport calculations. 
The $x\rightarrow 0$ limit of the above expression requires some care because of the appearance of a double pole. 
This is resolved by integrating by parts and one obtains:
\begin{equation}
\frac{T}{Z(T,\omega=0)}\,=\,T+\frac{\lambda}{2\pi} (\pi T)^\nu \,\int dy\, \frac{g^\prime(y)}{y}
\end{equation}
so that, with the notation introduced in the text:
\begin{equation}
\frac{T}{Z(T)}\,=\,T+\frac{\lambda (\pi T)^\nu}{\prefZ}\,\,\,,\,\,\,\frac{1}{\prefZ}=\frac{1}{2\pi}\,\int dy\, \frac{g^\prime(y)}{y}
\end{equation}

These expressions are valid for $\nu<1$. The Planckian case $\nu=1$ requires a slightly different analysis because in  
Eq.~(\ref{eq:resigma_KK}) the integral would diverge logarithmically at large frequency. 
Hence, a cutoff $\Lambda$ must be kept. The low-temperature behavior of the zero-frequency 
limit of $Z(T)$  can be again obtained from an integration by part (up to the cutoff).
We obtain: 
\begin{equation}
    1-1/Z(T) = -\frac{\lambda}{2}\left[P\int_{-\Lambda/T}^{+\Lambda/T} dy\,\frac{g^\prime(y)}{y} - \left. \frac{g(y)}{y} \right \rvert^{\Lambda/T}_{-\Lambda/T} \right]
\end{equation}
At large $y$, $g^\prime$ tends to a constant in the Planckian case 
$g^\prime(y\sim \pm \infty)\rightarrow  g^\prime(\pm\infty)$, so that the low temperature behavior is:
\begin{equation}
    \frac{1}{Z(T)}\,=\,1+\frac{\lambda}{2} \left[g^\prime(+\infty)-g^\prime(-\infty)\right]\,
    \ln\left(\zeta\frac{\Lambda}{T}\right) 
\end{equation}
with $\zeta\simeq 0.28$ a numerically determined constant.  
Hence, using the notation in the main text 
(and absorbing $\zeta$ in the definition of the cutoff $\Lambda$), we obtain: 
$2\pi/\prefZ = \left[g^\prime(+\infty)-g^\prime(-\infty)\right]$ 
($=1$ for any $\alpha$). In the Planckian case, the high-energy cutoff does not entirely disappear from the expression of $Z$, but we note that the prefactor 
of the $\ln T$ term 
depends only on the inelastic coupling constant and not on the cutoff.

\section{Temperature dependence of the Seebeck coefficient}
\label{app:T}
\begin{figure}
 \begin{center}
   \includegraphics[width=0.7\columnwidth,keepaspectratio]{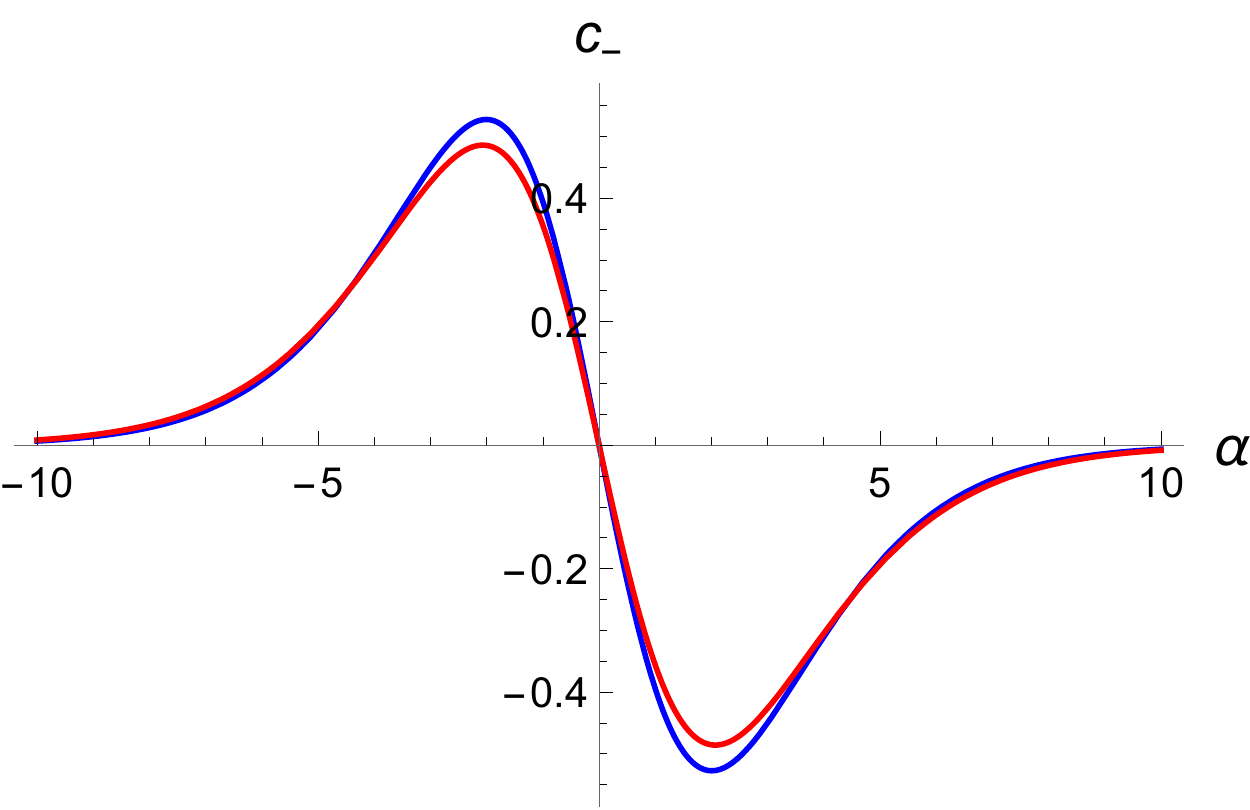}
   \end{center}
   \caption{Coefficient $c_{-}$ vs. $\alpha$ for $\nu=0.5$ (blue) and $\nu=1$ (red). 
   \label{fig:cminus}
   }
 \end{figure}

\begin{figure}
 \begin{center}
   \includegraphics[width=0.65\columnwidth,keepaspectratio]{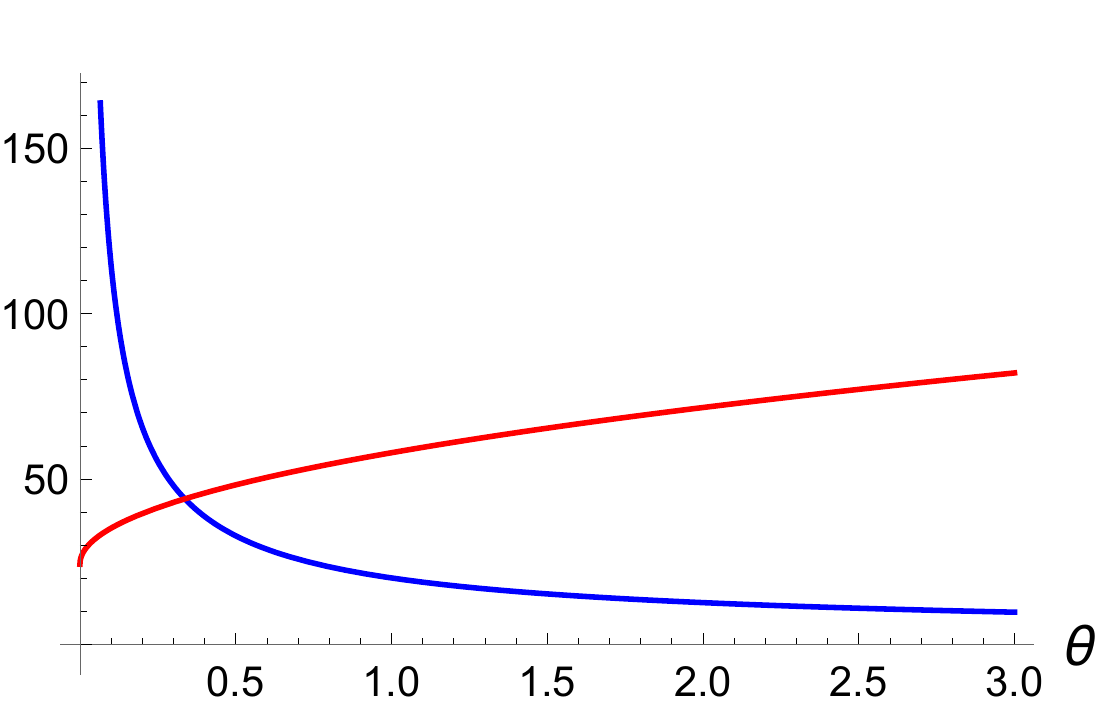}
   \end{center}
   \caption{$\ts^\nu F_2(\ts)/F_1(\ts)$ vs. $\ts$ (red) and 
   $F_0(\ts)/\ts^\nu F_1(\ts)$ vs. $\ts$ (blue) for $\nu=0.5$ and $\alpha=0.3$. 
   \label{fig:Fratio}
   }
 \end{figure}

Setting $\ts=T/T^*$ and $\eta=\gamel\php/\ph$, expressions Eqs.~(\ref{eq:I1},\ref{eq:I0}) can be written as: 
\begin{eqnarray}
I_1(\ts)&=& F_1(\ts) + \eta\left[\frac{T^*}{\gamel}\,\ts + \frac{1}{\prefZ} \ts^\nu \right]\,F_2(\ts)
\label{eq:I1theta}
\\
I_0(\ts)&=& F_0(\ts) + \eta\left[\frac{T^*}{\gamel}\,\ts + \frac{1}{\prefZ} \ts^\nu \right]\,F_1(\ts)
\label{eq:I0theta}
\end{eqnarray}
with:
\begin{equation}
F_n(\ts)\,=\,\langle \frac{x^n}{1+\ts^\nu g(x)} \rangle
\end{equation}
The low-temperature expansion ($\ts\rightarrow 0$) of the functions $F_n$ reads:
\begin{eqnarray}
F_0(\ts)&=&1- \braket{g}\,\ts^\nu + \braket{g^2}\,\theta^{2\nu}+\cdots\nonumber\\
F_1(\ts)&=& - \braket{xg}\,\ts^\nu + \braket{x g^2}\,\theta^{2\nu}+\cdots\nonumber\\
F_2(\ts)&=&\frac{\pi^2}{3}- \braket{x^2 g}\,\ts^\nu + 
\braket{x^2 g^2}\,\theta^{2\nu}+\cdots\nonumber\\
\end{eqnarray}
The key difference in the behavior of $F_n$ is that $F_1\sim c_{-}\ts^\nu$ ($c_{-}=\braket{xg}$) 
while $F_{0,2}\sim$const. 
As a result, $\ts^\nu F_{2}$ and $F_1$ have the same temperature dependence 
in this regime and hence both contribute to $I_1$ and to the thermopower. 
In contrast, in $I_0$, the term $\ts^\nu F_1\sim \ts^{2\nu}$ 
can be neglected in comparison to $F_0$ in both the conductivity 
and the Seebeck coefficient, which can be hence written as:
\begin{equation}\nonumber
S(T)\,=\,-\frac{k_B}{e}\,\left[
\eta\left(\frac{1}{\prefZ}\theta^\nu+\frac{T^*}{\gamma}\theta\right)\frac{F_2(\ts)}{F_0(\ts)} 
+ \frac{F_1(\ts)}{F_0(\ts)}\right].
    \label{eq:crossover}
\end{equation}
$F_1$ is much smaller in magnitude than $\ts^\nu F_2$, as shown 
on Fig.~\ref{fig:Fratio}. Hence, the odd-frequency contribution $F_1$ in $I_1$ 
is comparable to the even one $\eta\ts^\nu F_2$ only when the parameter $\eta=\gamma\php/\ph$ is 
small (of order $10^{-2}$). This is however typically the case since the elastic scattering rate 
is usually much smaller than electronic energy scales (the elastic scattering rate 
is usually much smaller than the bandwidth). 
Hence, at low-$T$ the Seebeck coefficient reads for $\nu<1$:
\begin{equation}
    S \sim -\frac{k_B}{e} \left(\frac{T}{T^*}\right)^\nu 
    \left[\frac{\pi^2}{3\prefZ}\,\eta-c_{-} \right]
\end{equation}
The dependence of the coefficient $c_{-}$ on the asymmetry parameter $\alpha$ for the 
choice Eq.~(\ref{eq:cft_scaling}) of scaling functions is displayed on Fig.~\ref{fig:cminus}. 

In the high-temperature limit ($\ts\rightarrow\infty$) , all the functions 
$F_n$ have the same temperature dependence $\sim c_n/\ts^\nu$: 
\begin{equation}
   F_n(\ts) = \frac{1}{\ts^\nu}\,\braket{\frac{x^n}{g}} -  \frac{1}{\ts^{2\nu}}\,\braket{\frac{x^n}{g^2}} + \cdots
\end{equation}
Hence formally, at very high $T$, $S$ tends to a constant $=- c_2/c_1\,k_B/e$ 
in the presence of odd frequency scattering. However, the term involving $F_1$ in $I_0$ 
remains small for most temperatures of interest as shown on Fig.~\ref{fig:Fratio}. 
Hence, in practice, the relevant high-$T$ behaviour of $S$ is: 
\begin{equation}
    S \sim -\frac{k_B}{e}\left[ 
    \frac{c_2}{c_0}\frac{\php}{\ph}\,T + 
    \eta\frac{c_2}{c_0\prefZ}\left(\frac{T}{T^*}\right)^\nu\,+\,
    \frac{c_1}{c_0}\right].
\end{equation}

\begin{thebibliography}{71}
\bibitem [{not({\natexlab{a}})}]{note_lowT}%
  \BibitemOpen
  \href@noop {} {} ({\natexlab{a}}),\ \bibinfo {note} {everywhere in the main
  text it is understood that temperature is low compared to the energy over
  which $\Phi$ varies, e.g. to the energy scale $\ph/\php$. Otherwise the
  low-$T$ expansion leading to (\ref{eq:I1},\ref{eq:I0}) is no longer valid,
  and the full Kubo formula detailed in the Appendix~\ref{app:kubo} has to be
  used.}\BibitemShut {Stop}%
\bibitem [{not({\natexlab{b}})}]{note_Z}%
  \BibitemOpen
  \href@noop {} {} ({\natexlab{b}}),\ \bibinfo {note} {$Z(T)$ is finite at
  $T\neq 0$ even when the low-frequency behaviour of the self-energy is
  singular at $T=0$.}\BibitemShut {Stop}%
\end{thebibliography}
%

\end{document}